\documentclass[12pt,a4paper]{article}
\pdfoutput=1

\usepackage[a4paper]{geometry}
\usepackage{amsmath}
\usepackage{amsfonts}
\usepackage{amssymb}
\usepackage[nosort]{cite}
\usepackage[backref=page]{hyperref}
\usepackage{graphicx}
\usepackage[margin=20pt,font=small,labelfont=bf]{caption}
\usepackage{multicol}

\newcommand{\eq}[1]{\begin{equation}
                     \begin{split} #1 \end{split}
                     \end{equation}}
\newcommand{\ov}{\overline}
\newcommand{\op}{\hspace{1pt}}

\allowdisplaybreaks[2]
\numberwithin{equation}{section}

\newcommand{\botxt}[1]{\begin{center}\fboxsep2mm
           \framebox{\begin{minipage}[t]{0.87\textwidth}#1 \end{minipage}}
           \end{center}}

\begin{document}

\vspace*{1.5cm}

\begin{center}
{\LARGE
Moduli stabilization in type IIB \\ orientifolds at $h^{2,1}=50$\\
}
\end{center}

\vspace{0.7cm}

\begin{center}
Konstantinos Tsagkaris, 
  Erik Plauschinn
\end{center}

\vspace{0.7cm}

\begin{center} 
\textit{
Institute for Theoretical Physics, Utrecht University \\
Princetonplein 5, 3584CC Utrecht \\
The Netherlands \\
}
\end{center} 

\vspace{2cm}

%%%%%%%%%%%%%%%%%%%%%%%%%%%%%%%%%%%%%%%%%%%%%%%
%%%%%%%%%%%%%%%%%%%%%%%%%%%%%%%%%%%%%%%%%%%%%%%
%%%%%%%%%%%%%%%%%%%%%%%%%%%%%%%%%%%%%%%%%%%%%%%
%%%%%%%%%%%%%%%%%%%%%%%%%%%%%%%%%%%%%%%%%%%%%%%

\begin{abstract}
\noindent
We study moduli stabilization in Calabi-Yau orientifold compactifications of type IIB 
string theory with O3- and O7-planes. We consider a Calabi-Yau three-fold
with Hodge number $h^{2,1}=50$  
and stabilize all axio-dilaton and complex-structure moduli 
by
three-form fluxes.
This is a challenging task, especially for large 
moduli-space dimensions. 
To address this question
we develop an algorithm to generate $10^5$ flux vacua 
with small flux number $N_{\rm flux}$.
Based on recent results by 
Crin\`o et al.~we estimate the bound imposed by 
the tadpole-cancellation condition as 
$N_{\rm flux}\leq \mathcal O(10^3)$, however,
the smallest flux number we obtain in our search is of order $N_{\rm flux}=\mathcal O(10^{4})$.
This implies, in particular, that for all solutions to the F-term equations 
in our data set  the tadpole conjecture is satisfied. 
\end{abstract}

%%%%%%%%%%%%%%%%%%%%%%%%%%%%%%%%%%%%%%%%%%%%%%%
%%%%%%%%%%%%%%%%%%%%%%%%%%%%%%%%%%%%%%%%%%%%%%%
%%%%%%%%%%%%%%%%%%%%%%%%%%%%%%%%%%%%%%%%%%%%%%%
%%%%%%%%%%%%%%%%%%%%%%%%%%%%%%%%%%%%%%%%%%%%%%%

\clearpage

%%%%%%%%%%%%%%%%%%%%%%%%%%%%%%%%%%%%%%%%%%%%%%%
%%%%%%%%%%%%%%%%%%%%%%%%%%%%%%%%%%%%%%%%%%%%%%%
%%%%%%%%%%%%%%%%%%%%%%%%%%%%%%%%%%%%%%%%%%%%%%%
%%%%%%%%%%%%%%%%%%%%%%%%%%%%%%%%%%%%%%%%%%%%%%%
%%%%%%%%%%%%%%%%%%%%%%%%%%%%%%%%%%%%%%%%%%%%%%%
%%%%%%%%%%%%%%%%%%%%%%%%%%%%%%%%%%%%%%%%%%%%%%%
%%%%%%%%%%%%%%%%%%%%%%%%%%%%%%%%%%%%%%%%%%%%%%%
%%%%%%%%%%%%%%%%%%%%%%%%%%%%%%%%%%%%%%%%%%%%%%%
%%%%%%%%%%%%%%%%%%%%%%%%%%%%%%%%%%%%%%%%%%%%%%%
%%%%%%%%%%%%%%%%%%%%%%%%%%%%%%%%%%%%%%%%%%%%%%%
%%%%%%%%%%%%%%%%%%%%%%%%%%%%%%%%%%%%%%%%%%%%%%%
%%%%%%%%%%%%%%%%%%%%%%%%%%%%%%%%%%%%%%%%%%%%%%%

\section{Introduction}

String theory is a consistent theory of quantum gravity including gauge interactions.
The theory is defined in ten space-time dimensions --- and the connection to physics in four dimensions 
is typically made by compactifying  on Calabi-Yau three-folds. 
The resulting effective theory is determined largely by the 
geometry of the compact space, in particular,  deformations that 
preserve the Calabi-Yau condition 
 correspond to massless scalar
fields. However, such fields are in conflict with experimental observations. 
One way to resolve this conflict is to include 
fluxes which can give masses to the moduli. 
Well-understood settings for this procedure are orientifold compactifications of type IIB string theory 
with O3- and O7-planes, where  
three-form fluxes  generate a  potential 
for the axio-dilaton and complex-structure moduli \cite{Giddings:2001yu}.
Stabilizing moduli in this way is the first step 
in the KKLT \cite{Kachru:2003aw} and Large Volume Scenarios \cite{Balasubramanian:2005zx}.

One may expect that a generic choice of fluxes will stabilize all 
axio-dilaton and complex-structure moduli.
This expectation has been challenged recently.
Fluxes are restricted by the geometry and topology of the 
compact space, more concretely,  the  flux number $N_{\rm flux}$ 
is bounded through the tadpole-cancellation condition. 
In \cite{Bena:2018fqc} it has  been argued that 
stabilizing moduli near a conifold locus requires large fluxes  which are 
incompatible with the tadpole-can\-cel\-la\-tion condition
and in \cite{Junghans:2022exo,Gao:2022fdi,Junghans:2022kxg} 
it has been discussed that for control of perturbative 
corrections in the Large Volume Sce\-na\-rio large flux numbers 
(likely exceeding the tadpole bound)
are needed.
In  \cite{Betzler:2019kon} we showed that 
the tadpole-cancellation condition can force moduli to be stabilized
in a perturbatively poorly-controlled regime
and in \cite{Braun:2020jrx} its is argued that 
 stabilizing all moduli in M-theory can be
in tension with the tadpole condition.
Taking these arguments one step further, 
the authors of \cite{Bena:2020xrh} made the \textit{tadpole conjecture}
which implies that stabilizing a large number of moduli by 
fluxes is not possible.

The purpose of the present work is to investigate the tadpole conjecture 
for a concrete setting. We consider a Calabi-Yau three-fold
with 50 complex-structure moduli for which the 
tadpole conjecture is applicable. 
All axio-dilaton and com\-plex-structure moduli 
are stabilized by fluxes in the large-complex-struc\-ture limit. 
This is a challenging task --- and we have
developed an algorithm
that allows us to construct a large number of flux vacua with a small flux number $N_{\rm flux}$.
Based on  results of \cite{Crino:2022zjk} we estimate a bound on $N_{\rm flux}$ from the tadpole-cancellation 
condition as $N_{\rm flux}\leq \mathcal O(10^3)$
and compare with our solutions. 
We find that the vacua in our data set 
satisfy $N_{\rm flux}\geq \mathcal O(10^{4})$ and therefore
exceed this bound. 
In particular, for the vacua we obtain the tadpole conjecture is satisfied.

This paper is organized as follows: in section~\ref{sec_prelim} we 
briefly review moduli stabilization for type IIB orientifolds. 
In section~\ref{sec_tadpole} we discuss the tadpole conjecture in some detail,
in section~\ref{sec_example} we introduce a concrete setting for studying 
moduli stabilization,
and in section~\ref{sec_results} we present and discuss our results.

%%%%%%%%%%%%%%%%%%%%%%%%%%%%%%%%%%%%%%%%%%%%%%%
%%%%%%%%%%%%%%%%%%%%%%%%%%%%%%%%%%%%%%%%%%%%%%%
%%%%%%%%%%%%%%%%%%%%%%%%%%%%%%%%%%%%%%%%%%%%%%%
%%%%%%%%%%%%%%%%%%%%%%%%%%%%%%%%%%%%%%%%%%%%%%%
%%%%%%%%%%%%%%%%%%%%%%%%%%%%%%%%%%%%%%%%%%%%%%%
%%%%%%%%%%%%%%%%%%%%%%%%%%%%%%%%%%%%%%%%%%%%%%%
%%%%%%%%%%%%%%%%%%%%%%%%%%%%%%%%%%%%%%%%%%%%%%%
%%%%%%%%%%%%%%%%%%%%%%%%%%%%%%%%%%%%%%%%%%%%%%%
%%%%%%%%%%%%%%%%%%%%%%%%%%%%%%%%%%%%%%%%%%%%%%%
%%%%%%%%%%%%%%%%%%%%%%%%%%%%%%%%%%%%%%%%%%%%%%%

\section{Moduli stabilization}
\label{sec_prelim}

In this section we  introduce the setting for our subsequent discussion. 
We focus on the material necessary for moduli stabilization in the large-complex-structure 
regime and refer for instance to \cite{Blumenhagen:2006ci} 
for a more detailed introduction to this topic.

%%%%%%%%%%%%%%%%%%%%%%%%%%%%%%%%%%%%%%%%%%%%%%%
%%%%%%%%%%%%%%%%%%%%%%%%%%%%%%%%%%%%%%%%%%%%%%%

\subsubsection*{Moduli}

We consider orientifold compactifications of type IIB string theory on Calabi-Yau three-folds $\mathcal X$
with O3- and O7-planes. The orientifold projection splits the cohomologies 
of   $\mathcal X$ into even and odd eigenspaces 
$H^{p,q}_{\pm}(\mathcal X)$, whose dimensions will be denoted by $h^{p,q}_{\pm}$.
The effective four-dimensional theory obtained after compactification 
contains massless scalar fields, in particular,
the axio-dilaton $\tau$, $h^{2,1}_-$ complex-structure moduli $z^i$, 
 $h^{1,1}_+$ K\"ahler moduli $T_a$, and  $h^{1,1}_-$ moduli $G_{\hat a}$.
We parametrize the first two as
\eq{
\tau = c+ i \op s \,, 
\hspace{70pt}
z^i = u^i + i \op v^i\,, 
\hspace{30pt}
i=1,\ldots, h^{2,1}_-\,,
}
and the physical region of the dilaton is characterized by $s>0$.
The K\"ahler potential for these  fields  is given by 
\eq{
\label{kpot}
  \mathcal K =  - \log\bigl[ -i(\tau-\bar \tau) \bigr] - \log \left[ +i\int_{\mathcal X} \Omega \wedge \bar \Omega \right]
  - 2\log \mathcal V\,,
}
where $\Omega$ denotes the holomorphic three-form of $\mathcal X$ 
which depends on the complex-structure moduli $z^i$
and $\mathcal V$ denotes the volume of $\mathcal X$  which 
depends on the K\"ahler moduli $T_a$ and on the moduli $G_{\hat a}$.

%%%%%%%%%%%%%%%%%%%%%%%%%%%%%%%%%%%%%%%%%%%%%%%
%%%%%%%%%%%%%%%%%%%%%%%%%%%%%%%%%%%%%%%%%%%%%%%

\subsubsection*{Prepotential}

For the third cohomology of the Calabi-Yau three-fold $\mathcal X$ we can choose an integral 
symplectic basis $\{\alpha_I,\beta^I\}\in H^3_-(\mathcal X,\mathbb Z)$.
The holomorphic three-form is expanded in this basis in the following way
\eq{
  \Omega = X^I \alpha_I - \mathcal F_I \beta^I \,, \hspace{40pt} I = 0, \ldots, h^{2,1}_-\,,
}
where the periods $\mathcal F_{I}$ can be expressed using a prepotential 
$\mathcal F$ as $\mathcal F_I = \partial_I \mathcal F$
with $\partial_I = \partial/\partial X^I$. 
The complex-structure moduli $z^i$ are written in terms of the projective coordinates $X^I$ 
as $z^i = X^i/X^0$.
In this paper we are  interested in the large-complex-structure regime, for which 
the prepotential splits into a perturbative and a non-perturbative part 
\eq{
  \label{ppot_02}
  \mathcal F = \mathcal F_{\rm pert} + \mathcal F_{\rm inst} \,.
}
The perturbative part  takes the following form \cite{Hosono:1994av}
(we follow the discussion and conventions of \cite{Demirtas:2021nlu})
\eq{
\label{ppot}
\mathcal F_{\rm pert} = - \frac{1}{3!} \op\frac{\tilde\kappa_{ijk} X^i X^j X^k}{X^0} + 
\frac{1}{2!} \op a_{ij} X^i X^j + b_i X^i X^0 + 
\frac{1}{2!} \op c \op
(X^0)^2  \,,
}
where $\tilde\kappa_{ijk}$  are the triple intersection numbers 
of the mirror-dual three-fold $\tilde{\mathcal X}$ and the constants $a_{ij}$, $b_i$, and $c$ are
given by
\eq{
  a_{ij} = \frac{1}{2} \left\{ \begin{array}{@{}l@{\hspace{16pt}}l} \tilde\kappa_{iij} & i\geq j\,, \\ \tilde\kappa_{ijj} & i <j \,,
  \end{array}\right.
  \hspace{30pt} 
  b_i = \frac{1}{24} \int_{\tilde{\mathcal X}} c_2(\tilde{\mathcal X}) \wedge \tilde\beta_i\,,
  \hspace{30pt} 
  c= \frac{\zeta(3)\chi(\tilde{ \mathcal X})}{(2\pi i)^3}\,.
}
Here,
$c_2(\tilde{\mathcal X}) $ denotes the second Chern class of $\tilde{\mathcal X}$,
$\{\tilde\beta_i\}$ is a basis of $H^2(\tilde{\mathcal X},\mathbb Z)$ mirror-dual 
to the three-forms $\beta_i \in H_-^3(\mathcal X, \mathbb Z)$,
and $\chi(\tilde{ \mathcal X})$ is the Euler number of \raisebox{0pt}[0pt][0pt]{$\tilde{\mathcal X}$}.
The non-perturbative part of the prepotential originates from world-sheet instanton 
corrections in the mirror-dual theory. It takes the form
\eq{
  \label{ppot_03}
  \mathcal F_{\rm inst} = -\frac{1}{(2\pi  i)^3}(X^0)^2
  \sum_{\vec q} N_{\vec q} \: \mbox{Li}_3 \left( e^{2\pi i \op {q}_i X^i/X^0} \right),
}
where $\mbox{Li}_s$ denotes the polylogarithm, the vector $\vec q$ represents effective curve classes in 
$H_2(\tilde{\mathcal X},\mathbb Z)$, and $N_{\vec q}$ are the genus-zero Gopakumar-Vafa 
invariants of $\tilde{\mathcal X}$.

%%%%%%%%%%%%%%%%%%%%%%%%%%%%%%%%%%%%%%%%%%%%%%%
%%%%%%%%%%%%%%%%%%%%%%%%%%%%%%%%%%%%%%%%%%%%%%%

\subsubsection*{Fluxes}

In order to stabilize the axio-dilaton and complex-structure moduli we 
consider non-vanishing 
Neveu-Schwarz--Neveu-Schwarz (NS-NS) and Ramond-Ramond (R-R)
three-form fluxes 
$H_3$ and $F_3$. 
These fluxes are integer quantized and can be expanded as
\eq{
  H_3 = h^I \alpha_I - h_I \op\beta^I\,,
  \hspace{50pt}
  F_3 = f^I \alpha_I - f_I \op\beta^I\,,  
}
where $h^I,h_I,f^I,f_I\in\mathbb Z$.
They generate a scalar potential for the four-dimensional theory that 
can be computed from  the superpotential
\eq{
\label{wpot}
  W = \int_{\mathcal X} \Omega \wedge G_3 \,,
  \hspace{50pt}
  G_3 = F_3 -\tau \op H_3 \,.
}
Since the K\"ahler potential \eqref{kpot} we are considering is of no-scale type and
because the superpotential \eqref{wpot} does not depend
on the  moduli $T_a$ or  $G_{\hat a}$, the standard F-term scalar potential 
can be brought into the form
\eq{
\label{spot}
  V = e^{\mathcal K }  F_{\vphantom{\ov \beta}\alpha} \op G^{\alpha\ov \beta}\op \ov F_{\ov \beta} \,.
}
The K\"ahler potential $\mathcal K$ was shown in \eqref{kpot},
the F-terms are given by the K\"ahler-covariant derivative of $W$ as $F_{\alpha} = \partial_{\alpha} W + (\partial_{\alpha} \mathcal K) W$ with $\alpha=(\tau,z^i)$,
and $G^{\alpha\ov \beta}$ denotes the inverse of the K\"ahler metric
$G_{\alpha \ov \beta} = \partial_{\vphantom{\ov \beta}\alpha} \partial_{\ov \beta}\op \mathcal K$.

%%%%%%%%%%%%%%%%%%%%%%%%%%%%%%%%%%%%%%%%%%%%%%%
%%%%%%%%%%%%%%%%%%%%%%%%%%%%%%%%%%%%%%%%%%%%%%%

\subsubsection*{Tadpole-cancellation condition}

The fixed loci of the orientifold projection give rise to orientifold planes.
These objects are charged under the R-R gauge potentials and 
therefore contribute to the corresponding Bianchi identities as sources. 
To solve these identities one typically has to introduce D-branes, which for our setting are  D3- and D7-branes. 
Integrating the Bianchi identities leads to the tadpole-cancellation 
conditions, and relevant for our discussion is the D3-brane tadpole given by (for  details 
on the derivation see for instance \cite{Plauschinn:2008yd})
\eq{
\label{tadpole_d3}
0=N_{\rm flux} +  2\op N_{{\rm D}3}  + Q_{\rm D3} \,,
}
where we defined
\begin{align}
  \label{tad_03}
  N_{\rm flux} &= \int_{\mathcal X} F_3 \wedge H_3  \,,
\\
\label{tadpole_d3b}
Q_{\rm D3} &= -  \frac{N_{{\rm O}3}}{2} 
  -\sum_{{\rm D}7_{\mathsf i}}\left[ \int_{\Gamma_{{\rm D}7_{\mathsf i}}}
  \hspace{-10pt}
    \mbox{tr}\left[ \mathsf F^2_{{\rm D}7_{\mathsf i}}\right] 
  + N_{{\rm D}7_{\mathsf i}}\op\frac{\chi( \Gamma_{{\rm D}7_{\mathsf i}}\bigr)}{12} 
  \right]
  -\sum_{{\rm O}7_{\mathsf j}} \frac{\chi\bigl( \Gamma_{{\rm O}7_{\mathsf j}}\bigr)}{6} \,.
\end{align}
The flux number $N_{\rm flux}$ is non-negative in our conventions, 
$N_{{\rm D}7_{\mathsf i}}$ denotes the number of D7-branes in a stack labelled by $\mathsf i$
wrapping a four-cycle $\Gamma_{{\rm D}7_{\mathsf i}}$ in $\mathcal X$,
and $N_{{\rm D}3}$ is the total number of D3-branes. Both of these numbers are counted 
without the orientifold images.
Furthermore, $\mathsf F_{{\rm D}7_{\mathsf i}}$ is the open-string gauge flux for a stack $\mathsf i$, 
$N_{{\rm O}3}$ is the total number of O3-planes and $\chi(\Gamma)$ denotes 
the Euler number of the cycle $\Gamma$.

%%%%%%%%%%%%%%%%%%%%%%%%%%%%%%%%%%%%%%%%%%%%%%%
%%%%%%%%%%%%%%%%%%%%%%%%%%%%%%%%%%%%%%%%%%%%%%%

\subsubsection*{Minima}

We are interested in global minima of the scalar potential \eqref{spot}.
These are given by vanishing F-terms $F_{\alpha}=0$
and to implement these conditions we expand the 
three-form flux $G_3$ appearing in \eqref{wpot}
in the integral symplectic basis $\{\alpha_I,\beta^I\}$ 
as 
\eq{
  G_3 = m^I \alpha_I - e_I \beta^I \,.
}
Next, we  denote the derivative of the periods by
$\mathcal{F}_{IJ} =\partial_I\mathcal{F}_J$ and  define the complex matrix 
\begin{equation}
    \mathcal{N}_{IJ} = \ov{\mathcal{F}}_{IJ} + 2i \,\frac{(\operatorname{Im}\mathcal{F})_{IM}X^M (\operatorname{Im}\mathcal{F})_{JN}X^N}{X^R (\operatorname{Im} \mathcal{F})_{RS} X^S}\,,
\end{equation}
with $I,J,\ldots=0,\ldots, h^{2,1}_-$. The requirement of vanishing F-terms, that is the minimum conditions, can then be expressed
as the following complex-valued matrix equation
\eq{
  \label{minimum}
  e_I - \ov{\mathcal N}_{IJ}\op  m^J = 0 \,.
}

%%%%%%%%%%%%%%%%%%%%%%%%%%%%%%%%%%%%%%%%%%%%%%%
%%%%%%%%%%%%%%%%%%%%%%%%%%%%%%%%%%%%%%%%%%%%%%%

\subsubsection*{Stretched K\"ahler cone}

The prepotential \eqref{ppot_02} is valid in the large-complex-structure regime of the com\-plex-structure
moduli space. Away from that limit the instanton sum appearing in \eqref{ppot_03}
may not converge and  the large-complex-structure expansion may break down. 
The validity of the expansion can be characterized using a \textit{stretched K\"ahler cone}
\cite{Demirtas:2018akl} (see also \cite{Cicoli:2018tcq}). More concretely,
\begin{itemize}

\item the geometry of the complex-structure moduli space of $\mathcal X$ in the large-complex-structure
regime is mirror-dual to the geometry of the K\"ahler moduli space of $\tilde{\mathcal X}$
in the large-volume regime. For the latter one finds a cone structure 
and hence, by mirror symmetry, also the complex-structure moduli 
are restricted to lie in a cone. 

%%%%%%%%%%%%%%%%%%%%%
%%%%%%%%%%%%%%%%%%%%%
\begin{figure}
\centering
\includegraphics[width=150pt]{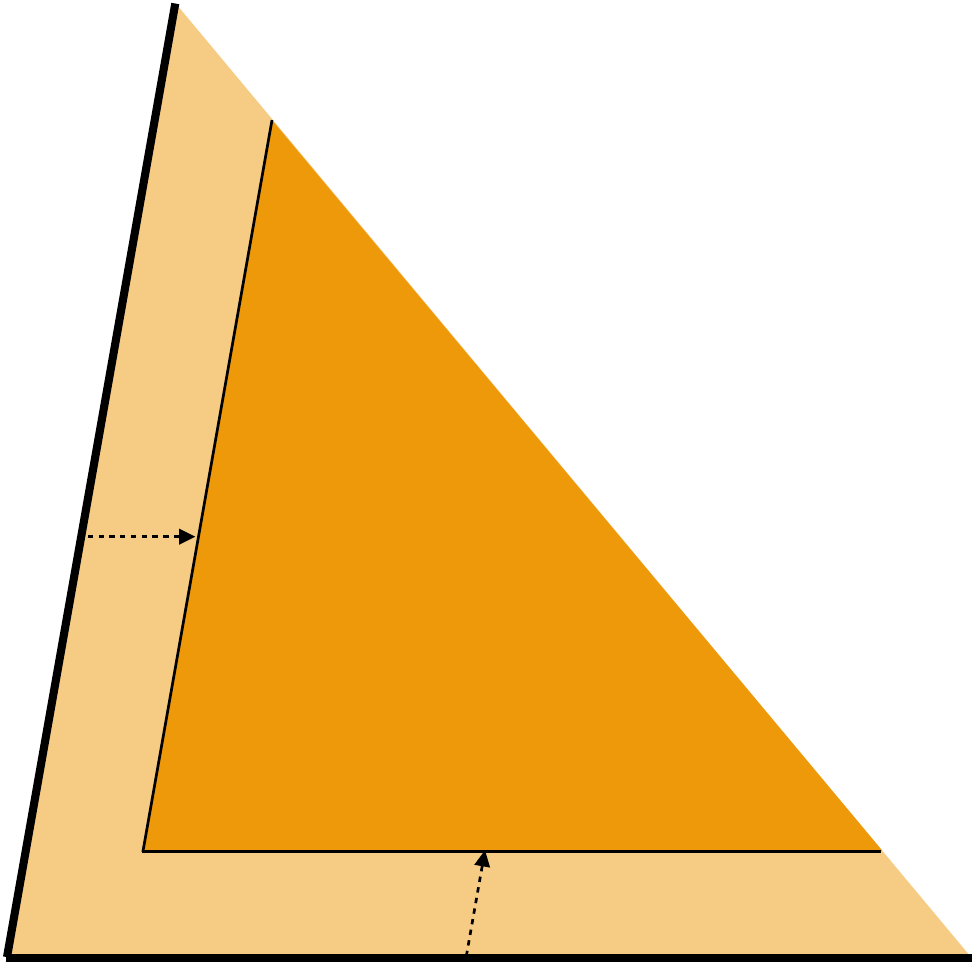}%
\begin{picture}(0,0)
\put(-74,7.5){\scriptsize$\mathsf c$}
\put(-130,69.5){\scriptsize$\mathsf c$}
\end{picture}
\caption{(Stretched) K\"ahler cones $\tilde{\mathcal K}_{\tilde{\mathcal X}}[0]$
and $\tilde{\mathcal K}_{\tilde{\mathcal X}}[\mathsf c]$
for a two-dimensional setting. 
The darker shaded region is the stretched K\"ahler cone with parameter~$\mathsf c$.
\label{fig_cone}}
\end{figure}
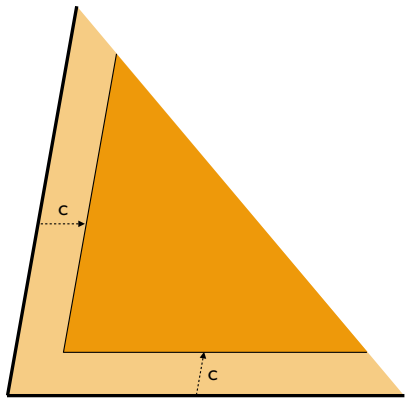
%%%%%%%%%%%%%%%%%%%%%
%%%%%%%%%%%%%%%%%%%%%

\item For the mirror-dual three-fold one can define a stretched K\"ahler cone in the following way
\cite{Demirtas:2018akl}
\eq{
  \label{skcone}
  \tilde{\mathcal K}_{\tilde{\mathcal X}}[\op \mathsf c\op] = \left\{ 
  \tilde J \in H^{1,1}(\tilde{\mathcal X},\mathbb R) : 
  \textrm{vol}(W) \geq \mathsf c \quad\forall \,W \in \mathcal W\right\} ,
}
where $\tilde J$ denotes the K\"ahler form on $\tilde{\mathcal X}$, $\mathcal W$ are all subvarieties (curves, divisors, $\tilde{\mathcal X}$) of 
the dual three-fold $\tilde{\mathcal X}$,
and $\mathsf c$ is a constant in  appropriate units (see figure~\ref{fig_cone} 
for an illustration). 
The ordinary K\"ahler cone is given by $\tilde{\mathcal K}_{\tilde{\mathcal X}}[0]$.

\item Coming back to the prepotential \eqref{ppot_02}, the sum of world-sheet instanton corrections \eqref{ppot_03}
converges when the volumes of curves are sufficiently large. 
Hence, in order to trust the large-complex-structure expansion, 
the complex-structure moduli should be stabilized inside 
a stretched K\"ahler cone
 with parameter 
$\mathsf c>0$. We come back to this point below.

\end{itemize}

%%%%%%%%%%%%%%%%%%%%%%%%%%%%%%%%%%%%%%%%%%%%%%%
%%%%%%%%%%%%%%%%%%%%%%%%%%%%%%%%%%%%%%%%%%%%%%%
%%%%%%%%%%%%%%%%%%%%%%%%%%%%%%%%%%%%%%%%%%%%%%%
%%%%%%%%%%%%%%%%%%%%%%%%%%%%%%%%%%%%%%%%%%%%%%%
%%%%%%%%%%%%%%%%%%%%%%%%%%%%%%%%%%%%%%%%%%%%%%%
%%%%%%%%%%%%%%%%%%%%%%%%%%%%%%%%%%%%%%%%%%%%%%%
%%%%%%%%%%%%%%%%%%%%%%%%%%%%%%%%%%%%%%%%%%%%%%%
%%%%%%%%%%%%%%%%%%%%%%%%%%%%%%%%%%%%%%%%%%%%%%%
%%%%%%%%%%%%%%%%%%%%%%%%%%%%%%%%%%%%%%%%%%%%%%%
%%%%%%%%%%%%%%%%%%%%%%%%%%%%%%%%%%%%%%%%%%%%%%%

\section{Tadpole contribution of fluxes}
\label{sec_tadpole}

The NS-NS and R-R three-form fluxes $H_3$ and $F_3$ generate a potential in the 
effective four-dimensional theory that can stabilize the axio-dilaton and complex-structure 
moduli. However, $H_3$ and $F_3$ cannot be chosen arbitrarily but are restricted by the tadpole cancellation 
condition  \eqref{tadpole_d3}.
In this section we discuss the interplay between these two questions and motivate 
the setting for our subsequent analysis.

%%%%%%%%%%%%%%%%%%%%%%%%%%%%%%%%%%%%%%%%%%%%%%%
%%%%%%%%%%%%%%%%%%%%%%%%%%%%%%%%%%%%%%%%%%%%%%%

\subsubsection*{The tadpole conjecture}

Let us denote the number of axio-dilaton and complex-structure moduli 
which are stabilized through the flux superpotential \eqref{wpot} by 
$n_{\rm stab}$. 
In \cite{Bena:2020xrh} the authors conjectured that for $n_{\rm stab}\gg 1$
the flux number $N_{\rm flux}$ shown in \eqref{tad_03} grows at least linearly with $n_{\rm stab}$.
This conjecture has been called the \textit{tadpole conjecture} and several versions thereof exist. 
In this paper we are interested in stabilizing 
all axio-dilaton and complex-structure moduli in type IIB orientifold compactifications by fluxes.
In this case
the conjecture takes the form
\eq{
  \label{tad_conj}
   N_{\rm flux}  > 2\op\alpha\op (h^{2,1}_-+1)
  \hspace{50pt} \mbox{for}\quad h^{2,1}_-\gg 1\,,
}
and
in the refined version the constant $\alpha$ is conjectured to be $\alpha=1/3$. 
We briefly summarize the current status of this conjecture:
\begin{itemize}

\item 
The tadpole conjecture has been verified 
for F-theory compactifications on $K3\times K3$  in  \cite{Bena:2020xrh,Bena:2021wyr},
for type IIB compactifications in the large complex-structure regime in 
\cite{Plauschinn:2021hkp}, 
for  F-theory compactifications on Calabi-Yau four-folds with a weak Fano base
in \cite{Bena:2021qty}, and for F-theory compactifications in asymptotic regimes in 
\cite{Grana:2022dfw}.

\item A counter-example for the tadpole conjecture was proposed in 
\cite{Marchesano:2021gyv}, but was opposed in 
 \cite{Lust:2021xds} and \cite{Grimm:2021ckh}.
In particular, currently no concretely worked-out  model that violates the tadpole conjecture 
is known to exist. 
 
\item The tadpole conjecture applies to smooth compactifications but could be
violated for spaces that contain singularities. This was already noted in 
\cite{Bena:2020xrh} and has been emphasized for instance 
in \cite{Gao:2022fdi}.

\end{itemize}

%%%%%%%%%%%%%%%%%%%%%%%%%%%%%%%%%%%%%%%%%%%%%%%
%%%%%%%%%%%%%%%%%%%%%%%%%%%%%%%%%%%%%%%%%%%%%%%

\subsubsection*{Contribution of orientifold planes and D7-branes}

The flux number $N_{\rm flux}$ is bounded from below by the requirement to be positive 
and  from above by the charge $Q_{\rm D3}$ 
in the tadpole cancellation condition \eqref{tadpole_d3}.
The contribution of orientifold planes appearing in $Q_{\rm D3}$ depends on the chosen orientifold projection
and it is difficult to give a precise estimate on how it varies with $h^{2,1}_-$.
For some classes of models complete classifications of the possible orientifold projections 
have been provided 
--- see for instance \cite{Lust:2006zh}
for a classification of orientifolds of $\mathbb T^6/\mathbb Z_M$ or
$\mathbb T^6/\mathbb Z_M\times \mathbb Z_N$
and \cite{Blumenhagen:2008zz} for a classification for del-Pezzo surfaces.
Furthermore,  databases for orientifold compactifications have been constructed recently:
in  \cite{Carta:2020ohw} orientifolds of CICYs have been classified,
in \cite{Altman:2021pyc} the authors extend their database
on triangulations for the Kreuzer-Skarke list \cite{Altman:2014bfa,Altman:2017vzk}
to include orientifold projections, 
and in  \cite{Crino:2022zjk} the authors systematically analyze 
the Kreuzer-Skarke database and classify orientifold projections. 
These analyses lead to the following bound on the 
D3-brane charge \eqref{tadpole_d3b} in type IIB orientifold compactifications
(see for instance equation (3.31) in  \cite{Crino:2022zjk}) 
\eq{
  \label{tadpole_bound}
  Q_{\rm D3} \geq - \bigl( 2 + h^{1,1}+ h^{2,1}  \bigr) \,.
}
Note that  this bound is obtained for configurations in which D7-branes are placed 
on top of O7-planes and hence charges are cancelled locally. 
However, 
when considering D7-branes away from the orientifold locus 
(i.e.~non-local cancellation of charges)
smaller
values for $ Q_{\rm D3} $ can be found  \cite{Collinucci:2008pf}.
From the database of \cite{Crino:2022zjk} we can infer the following 
bounds on $Q_{\rm D3}$ for local and non-local cancellation of D7-brane charges
\begin{align}
  \label{tadpole_bound_2}
  &Q_{\rm D3} \geq \left\{
  \begin{array}{rr}
  -48 & \mbox{local,}
  \\[4pt]
  -342 & \mbox{non-local,}
  \end{array}
  \right.
  \hspace{30pt}\mbox{for}\hspace{30pt}
  h^{1,1} = 5\,,\hspace{39.5pt} h^{2,1}_- = 50\,,
  \\[8pt]
  \label{tadpole_bound_3}
  &Q_{\rm D3} \geq \left\{
  \begin{array}{rr}
  -64 & \mbox{local,}
  \\[4pt]
  -722 & \mbox{non-local,}
  \end{array}
  \right.
  \hspace{30pt}\mbox{for}\hspace{30pt}
  h^{1,1} = 5\,,\quad 40\leq h^{2,1}_- \leq 60\,.
\end{align}
Since the dependence of the bounds on $Q_{\rm D3}$ may fluctuate with $h^{2,1}_-$, 
in  \eqref{tadpole_bound_3} we have broadened the search range to allow for 
variations in $h^{2,1}_-$. The values shown in \eqref{tadpole_bound_3} correspond to a model 
with $h^{2,1}_{\vphantom{-}}=h^{2,1}_-=57$.

%%%%%%%%%%%%%%%%%%%%%%%%%%%%%%%%%%%%%%%%%%%%%%%
%%%%%%%%%%%%%%%%%%%%%%%%%%%%%%%%%%%%%%%%%%%%%%%

\subsubsection*{Contribution of fluxes}

We note that a priori the bound shown in \eqref{tadpole_bound} is  compatible with 
the tadpole conjecture. However, the flux number 
$N_{\rm flux}$ may not saturate \eqref{tad_conj} but exceed it. 
In fact, for a toroidal orientifold compactification it was
observed in \cite{Betzler:2019kon}  that the flux number 
will generically diverge when moduli are stabilized near the boundary of moduli space 
while it will take its smallest values in the interior. 
In \cite{Plauschinn:2021hkp} this argument has been made for more general settings.

Motivated by the above works, we are interested in what values 
the flux number $N_{\rm flux}$ typically takes when many moduli are stabilized. 
To study this question we consider 
the large-complex-structure regime for which  the prepotential can easily be determined via mirror symmetry.
The validity of the large-complex-structure approximation  can be parametrized by 
the stretched K\"ahler-cone parameter $\mathsf c$ and 
we expect that for decreasing $\mathsf c$ the minimal $N_{\rm flux}$ will decrease 
--- while  at the same time the large-complex-structure expansion becomes less reliable. 
Similarly, the weak-string-coupling approximation is controlled by the dilaton (encoded in) $s$.
In view of the tadpole conjecture we are therefore interested in the question:
\botxt{
For a given bound on the K\"ahler-cone parameter $\mathsf c$ and on the dilaton $s$, what is the smallest value of 
$N_{\rm flux}$ such that all axio-dilaton and complex-structure moduli are stabilized by fluxes?
}

%%%%%%%%%%%%%%%%%%%%%%%%%%%%%%%%%%%%%%%%%%%%%%%
%%%%%%%%%%%%%%%%%%%%%%%%%%%%%%%%%%%%%%%%%%%%%%%
%%%%%%%%%%%%%%%%%%%%%%%%%%%%%%%%%%%%%%%%%%%%%%%
%%%%%%%%%%%%%%%%%%%%%%%%%%%%%%%%%%%%%%%%%%%%%%%
%%%%%%%%%%%%%%%%%%%%%%%%%%%%%%%%%%%%%%%%%%%%%%%
%%%%%%%%%%%%%%%%%%%%%%%%%%%%%%%%%%%%%%%%%%%%%%%
%%%%%%%%%%%%%%%%%%%%%%%%%%%%%%%%%%%%%%%%%%%%%%%
%%%%%%%%%%%%%%%%%%%%%%%%%%%%%%%%%%%%%%%%%%%%%%%
%%%%%%%%%%%%%%%%%%%%%%%%%%%%%%%%%%%%%%%%%%%%%%%
%%%%%%%%%%%%%%%%%%%%%%%%%%%%%%%%%%%%%%%%%%%%%%%

\section{An example with $h^{2,1}_-=50$}
\label{sec_example}

In this section we introduce a
concrete setting for stabilizing 
 all axio-dilaton and complex-structure
moduli through NS-NS and R-R three-form fluxes.
We chose an example 
with $h^{2,1}_-=50$
for which the tadpole conjecture  is applicable.

%%%%%%%%%%%%%%%%%%%%%%%%%%%%%%%%%%%%%%%%%%%%%%%
%%%%%%%%%%%%%%%%%%%%%%%%%%%%%%%%%%%%%%%%%%%%%%%

\subsubsection*{The compactification space}

As discussed in section~\ref{sec_prelim}, we employ mirror symmetry to construct the
prepotential for the complex-structure moduli space
in the large-complex-structure regime. 
We use  \texttt{CYTools} \cite{Demirtas:2022hqf} to 
determine the data for the mirror-dual setting as follows:
\begin{itemize}

\item We have randomly chosen a polytope of the Kreuzer-Skarke 
database \cite{Kreuzer:2000xy}. Its normal form is shown in appendix~\ref{app_model}.
With the help of \texttt{CYTools}  we then perform a Delaunay triangulation using the command \texttt{triangulate()}
and  corresponding data is displayed  as well in appendix~\ref{app_model}.
The Calabi-Yau three-fold $\tilde{\mathcal X}$ is then obtained
by \texttt{get\_cy()}.

\item The Hodge numbers for the mirror three-fold are determined using the functions \texttt{h11} and 
\texttt{h12} as
\eq{
\label{model_001}
\tilde h^{1,1}= 50\,, \hspace{50pt} \tilde h^{2,1}= 5\,.
}
The perturbative part of the prepotential is specified by the triple-inter\-section numbers of $\tilde{\mathcal X}$ 
and the second Chern class, which are found using
\texttt{intersection\_numbers(in\_basis=true)} as well as using the command
 \texttt{second\_chern\_class(in\_basis=true)}.
The Euler number of the mirror-dual three-fold is computed from \eqref{model_001} as
\eq{
\chi(\tilde{\mathcal X}) = 90\,.
 }

\item The K\"ahler-cone conditions are encoded in a matrix $\mathcal M$, which  
can be computed using \texttt{CYTools} through
\texttt{toric\_kahler\_cone().hyperplanes()}.
A stretched K\"ahler cone with parameter $\mathsf c$ is specified by 
the conditions
\eq{
  \mathcal M \op \vec v \geq \vec{\mathsf c} \,,
}
where $\vec v$ is a vector with components $v^i = \mbox{Im}\op z^i$ and 
$\vec{\mathsf c}$ is a vector where each component takes the value $\mathsf c$. 

\item Finally, we assume that an orientifold projection can be chosen such that $h^{2,1}_-=h^{2,1}$, 
so that from \eqref{model_001} we have 
\eq{
h^{2,1}_- =50\,.
}
The tadpole cancellation condition \eqref{tadpole_d3}
relates the 
bounds on $Q_{\rm D3}$ shown in \eqref{tadpole_bound_2} and \eqref{tadpole_bound_3} for 
non-local cancellation of D7-branes charges 
to the following bound on the flux number 
\eq{
\label{flux_bound}
  N_{\rm flux} \leq \mathcal O(10^3) \,.
}

\end{itemize}

%%%%%%%%%%%%%%%%%%%%%%%%%%%%%%%%%%%%%%%%%%%%%%%
%%%%%%%%%%%%%%%%%%%%%%%%%%%%%%%%%%%%%%%%%%%%%%%

\subsubsection*{Estimating the validity of the large-complex-structure approximation}

The prepotential \eqref{ppot_02} can be separated into a perturbative and a non-perturbative contribution. 
The perturbative part \eqref{ppot}   is specified by the data discussed above, 
however, for the non-perturbative part \eqref{ppot_03} the Gopakumar-Vafa
invariants $N_{\vec q}$ are needed. 
For our setting we currently do not have access to that data, 
although this functionality is expected to be implemented in \texttt{CYTools} in the future. 
We therefore want to determine a criterion that characterizes when non-perturbative 
contributions to the prepotential can be neglected.

Gopakumar-Vafa invariants relevant for one particular 
point in complex-struc\-ture moduli space were kindly provided to us by 
J.~Moritz (see  \cite{Cornell}).
In the following we will denote this point by $\vec z_* = i\op \vec v_*$
and the vector $\vec v_*$ is shown in appendix~\ref{app_model}.
This point  is located at the tip of a stretched K\"ahler cone with parameter $\mathsf c=1$ 
and there are in total  $230$ effective curves with wrapping numbers $\vec q$ 
that satisfy $\vec q \cdot \vec v_*\leq 7$.
Next, when scaling the vector $\vec v_*$ with a parameter $\alpha>0$ we obtain a family with
\eq{
  \label{exp_800}
  z^i(\alpha)  = i\op \alpha\op z^i_*\,, \hspace{50pt}
  \mathsf c = \alpha\,.
}
For this family we
compute $\mathcal F_I =\partial_I \mathcal F$ for the perturbative 
and non-perturbative part of the prepotential and  define the ratios
\eq{
  r_I (\alpha)=  \left.\frac{\mathcal F_{{\rm inst}\op I}}{\mathcal F_{{\rm pert}\op I}}\right\rvert_{z^i(\alpha)}\,.
}
When $r_I=1$ for some $I$ the non-perturbative contribution is comparable to the perturbative 
part and instanton corrections cannot be neglected. 
We then determine numerically the largest $\alpha$ for which any ratio $r_I$ becomes one. We obtain
\eq{
\label{estimate_001}
 \max r_I(\alpha) = 1 \hspace{40pt}\Rightarrow\hspace{40pt}
 \alpha\simeq0.04
 \,.
}
We conclude from this analysis that along the ray $\vec{z}(\alpha) = \alpha\op \vec z_*$ 
the large-complex-structure approximation can at most be trusted 
for $\alpha\geq 0.04$ --- although it is very likely that the 
actual lower bound on $\alpha$ is larger. 
In the absence of complete information of the Gopakumar-Vafa invariants for our model, 
we  
estimate that  the large-complex-structure approximation requires
stretched K\"aher-cone parameters at least of the form
\eq{
  \mathsf c \geq 10^{-2} \,.
}

%%%%%%%%%%%%%%%%%%%%%%%%%%%%%%%%%%%%%%%%%%%%%%%
%%%%%%%%%%%%%%%%%%%%%%%%%%%%%%%%%%%%%%%%%%%%%%%

\subsubsection*{Constructing solutions}

In the following we want to solve the minimum conditions \eqref{minimum} 
computed from the perturbative 
prepotential $\mathcal F_{\rm pert}$ while ignoring the non-perturbative part. 
These conditions form a system of $102$ real coupled polynomial equations which (in practice) 
cannot be solved analytically. 
Even numerically it is extremely difficult to obtain solutions because search algorithms 
typically require a starting point near a minimum. Finding a suitable starting point by random 
sampling is highly unlikely, especially for high-dimensional moduli spaces.

However, as stated at the end of section~\ref{sec_tadpole}, 
we are interested in solutions with a small flux number $N_{\rm flux}$ 
for a given bound on the K\"ahler-cone parameter 
$\mathsf c$ such that all axio-dilaton and complex-structure moduli are 
stabilized. We developed an algorithm to construct such solutions with minimal $N_{\rm flux}$ 
is as follows:
\begin{itemize}

\item We first specify a stretched K\"ahler cone with parameter
$\mathsf c$. For each iteration of the algorithm $\mathsf c$ is chosen from a random 
uniform distribution in the range $\mathsf c_{(0)}\in[0.1,2]$.

\item Next, we randomly choose a point $\{ \tau^{\hphantom{i}}_{(0)}, z^i_{(0)}\}$ 
in axio-dilaton and complex-structure moduli 
space. This point is taken from a uniform distribution in the ranges
\eq{
  \label{sol_000}
s_{(0)}\in(0,10] \,, \hspace{35pt}
c_{(0)}\in(-0.5,+0.5]\,, \hspace{35pt}
u^i_{(0)}\in(-0.5,+0.5]\,.
}
The point  $v^i_{(0)}$ is obtained by determining  the tip of the stretched K\"ahler cone
\raisebox{0pt}[0pt][0pt]{$\tilde{\mathcal K}_{\tilde{\mathcal X}}[\mathsf c_{(0)}]$}.
Note that in order to illustrate the dependence of $N_{\rm flux}$ on $s={\rm Im}\op \tau$ we sample  
$s_{(0)}$ also in the strong-coupling regime;  we come back to this point below.

\item 
The matrix $\mathcal N$ appearing 
in the minimum conditions \eqref{minimum}
scales approximately linearly with  $v^i= {\rm Im}\op z^i$. 
At large complex structure, the fluxes $e_I$ therefore have to scale approximately as
$v^i \times m^J$. Hence, to minimize the flux number we are
interested in small values for $m^J$ for which we choose randomly  as
\eq{
  \label{sol_001}
  f_{(0)}^I,\op h_{(0)}^I\in\{-1,0,+1\}\,.
}

\item Inserting the values  $\{ \tau^{\hphantom{i}}_{(0)}, z^i_{(0)}\}$  and the fluxes
$\{f^I_{(0)},\op h^I_{(0)}\}$ into the minimum condition  \eqref{minimum}, we can easily solve 
for the fluxes $f_{(0)I}$ and $h_{(0)I}$. These are in general not integer, but we round them to integer values
denoted by $f_{(1)I},\op h_{(1)I}\in\mathbb Z$.

\item Next, we insert the integer-valued fluxes 
$\{f_{(0)}^I,\op h_{(0)}^I,\op f^{\vphantom{I}}_{(1)I},\op  h^{\vphantom{I}}_{(1)I}\}$
into the minimum condition.
We  solve \eqref{minimum} with \texttt{Python}'s
\texttt{scipy.optimize.root()}
and specify as starting point for the search algorithm 
\raisebox{0pt}[0pt][0pt]{$\{ \tau^{\vphantom{i}}_{(0)}, z^i_{(0)}\}$}. 
If for this choice of fluxes a solution 
\raisebox{0pt}[0pt][0pt]{$\{ \tau^{\hphantom{i}}_{(1)}, z^i_{(1)}\}$} is obtained,
we determine the Hessian of the scalar potential at this point. 
If the Hessian is of full rank, we  
add the solution to our data set. Note that these solutions can have a K\"ahler-cone parameter $\mathsf c_{(1)}$ 
that is smaller than the initialization bound $\mathsf c_{(0)}\geq 0.1$.

\end{itemize}
Using this strategy, we have generated $8.3\cdot 10^4$ solutions in the  large-complex-struc\-ture
regime that stabilized all axio-dilaton and complex-struc\-ture moduli. 
Let us  emphasize that our flux configurations are, 
on the one hand, non-generic among all possible flux configurations but, on the 
other hand,  generic for solutions in the large-complex-structure regime.
In particular, there can be (tuned) flux choices with a hierarchy 
different than the one described above which nevertheless stabilize moduli in the 
large-complex-structure regime with a small flux number.

%%%%%%%%%%%%%%%%%%%%%%%%%%%%%%%%%%%%%%%%%%%%%%%
%%%%%%%%%%%%%%%%%%%%%%%%%%%%%%%%%%%%%%%%%%%%%%%

\subsubsection*{Optimizing  solutions}

After having obtained solutions to the minimum conditions
in the large-complex-structure regime, we use this data 
as starting points for minimizing the flux number 
$N_{\rm flux}$. Our approach is as follows:
\begin{itemize}

\item We consider a flux configuration 
\raisebox{0pt}[0pt][0pt]{$\{ h_{(1)I}^{\vphantom{I}},\op h_{(1)}^I,\op f_{(1)I}^{\vphantom{I}},\op f_{(1)}^I\}$} 
with flux number $N_{{\rm flux}(1)}$ 
together 
with the corresponding minimum locus $\{ \tau^{\vphantom{i}}_{(1)},\op z^i_{(1)}\}$.

\item Next, we randomly change one of the flux quanta by one unit
and denote the modified fluxes by
\raisebox{0pt}[0pt][0pt]{$\{ h_{(2)I}^{\vphantom{I}},\op h_{(2)}^I,\op f_{(2)I}^{\vphantom{I}},\op f_{(2)}^I\}$}.
If through that change the corresponding flux number
$N_{{\rm flux}(2)}$ is smaller than $N_{{\rm flux}(1)}$, we numerically solve the minimum conditions for the new 
flux configuration with $\{ \tau^{\vphantom{i}}_{(1)},\op z^i_{(1)}\}$ as starting point. 

\item If a  solution 
\raisebox{0pt}[0pt][0pt]{$\{ \tau^{\vphantom{i}}_{(2)},\op z^i_{(2)}\}$} is found
that stabilizes all moduli 
(i.e.~the corresponding Hessian of the scalar potential is of full rank)
we repeat the algorithm with the new solution as starting point.

\end{itemize}
Through this mechanism we generated $1.7\cdot 10^4$ additional solutions that we
included in our data set.

%%%%%%%%%%%%%%%%%%%%%%%%%%%%%%%%%%%%%%%%%%%%%%%
%%%%%%%%%%%%%%%%%%%%%%%%%%%%%%%%%%%%%%%%%%%%%%%
%%%%%%%%%%%%%%%%%%%%%%%%%%%%%%%%%%%%%%%%%%%%%%%
%%%%%%%%%%%%%%%%%%%%%%%%%%%%%%%%%%%%%%%%%%%%%%%
%%%%%%%%%%%%%%%%%%%%%%%%%%%%%%%%%%%%%%%%%%%%%%%
%%%%%%%%%%%%%%%%%%%%%%%%%%%%%%%%%%%%%%%%%%%%%%%
%%%%%%%%%%%%%%%%%%%%%%%%%%%%%%%%%%%%%%%%%%%%%%%
%%%%%%%%%%%%%%%%%%%%%%%%%%%%%%%%%%%%%%%%%%%%%%%
%%%%%%%%%%%%%%%%%%%%%%%%%%%%%%%%%%%%%%%%%%%%%%%
%%%%%%%%%%%%%%%%%%%%%%%%%%%%%%%%%%%%%%%%%%%%%%%

\section{Results and discussion}
\label{sec_results}

In this section we present the data obtained from the analysis outlined in section~\ref{sec_example}.
This data, together with a corresponding \texttt{Mathematica} notebook explaining our conventions, can be found 
on the \href{https://arxiv.org/abs/2207.13721}{\texttt{arXiv} page} for this paper.
In the following we show how the flux number  $N_{\rm flux}$ depends on the K\"ahler-cone parameter $\mathsf c$
and on the string coupling encoded in $s$.
We furthermore discuss the implications of our findings and give an outlook for future work.

%%%%%%%%%%%%%%%%%%%%%%%%%%%%%%%%%%%%%%%%%%%%%%%
%%%%%%%%%%%%%%%%%%%%%%%%%%%%%%%%%%%%%%%%%%%%%%%

\subsubsection*{Numerical results}

In figures~\ref{fig_data_01} and \ref{fig_data_02} we have shown how, in our data, the minimal flux number
 depends on the K\"ahler cone parameter $\mathsf c$
 and on the string coupling  $s$.
We  only included data points with $N_{\rm flux}\leq 10^5$ in these plots, though
we  obtained many more with larger flux numbers.
We make the following observations:
\begin{itemize}

\item The minimal flux number  depends on 
the K\"ahler-cone parameter $\mathsf c$. 
For smaller $\mathsf c$ we
find a smaller $N_{\rm flux}$
(in agreement 
with \cite{Betzler:2019kon,Plauschinn:2021hkp}), however, for smaller $\mathsf c$
the large-complex-structure approximation becomes less reliable.
In figure~\ref{fig_data_01} we have shown $N_{\rm flux}(\mathsf c)$ 
for which we determined the bound 
\eq{
  \label{bound_01}
  N_{\rm flux}(\mathsf c)>
  2100\op \mathsf c + 880 \,.
}  
Note that for data points with $f^0\neq 0$ or $h^0\neq0$ (shown in blue in 
 figure~\ref{fig_data_01}) the bound on flux number scales as $N_{\rm flux }\sim \mathsf c^3$.

%%%%%%%%%%%%%%%%%%%%%
%%%%%%%%%%%%%%%%%%%%%
\begin{figure}[p]
\centering
\includegraphics[width=0.83\textwidth]{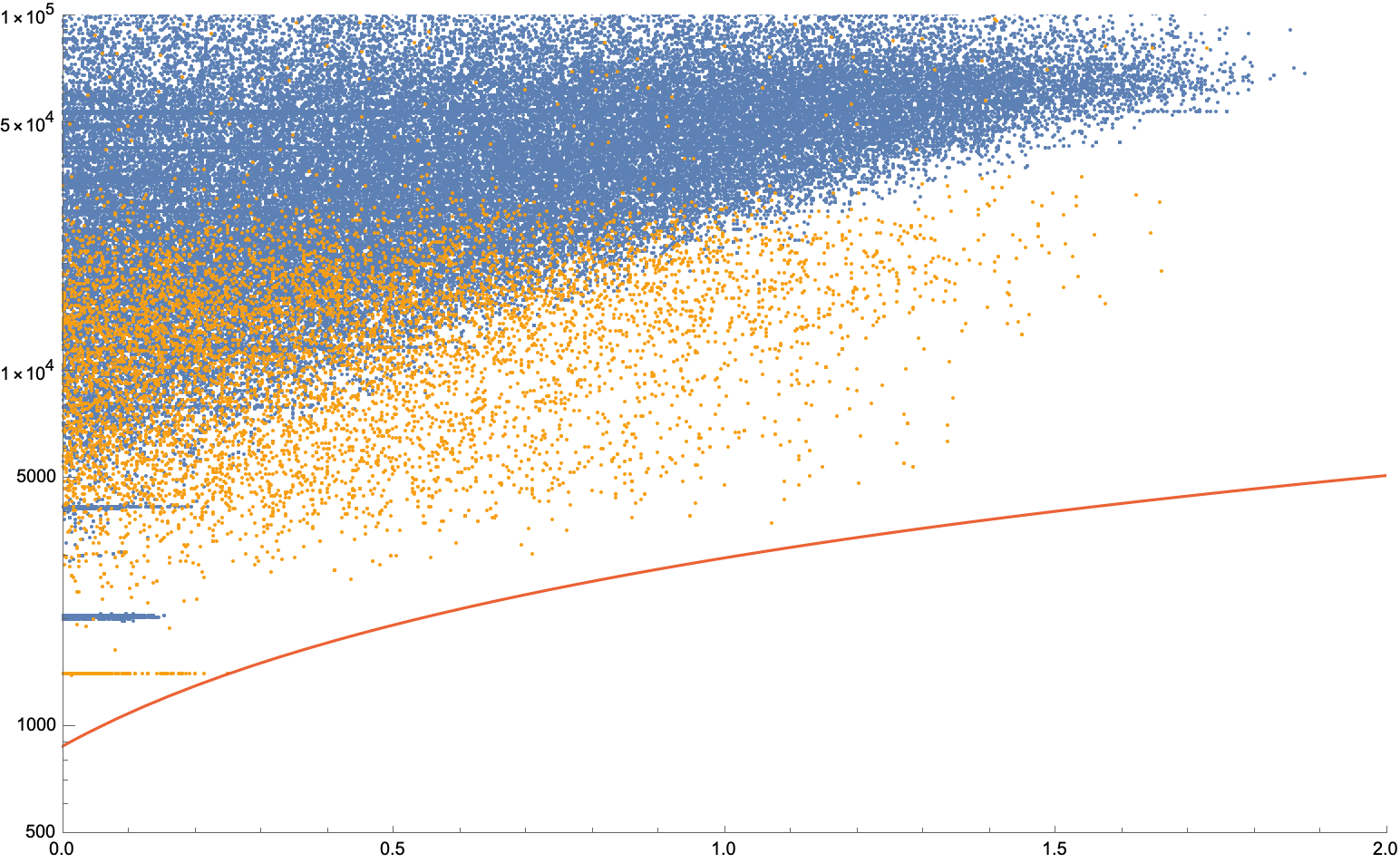}%
\begin{picture}(0,0)
\put(-352,218){\scriptsize$N_{\rm flux}$}
\put(2,4){\scriptsize$\mathsf c$}
\end{picture}
\caption{Dependence of $N_{\rm flux}$  on 
the K\"ahler-cone parameter $\mathsf c$.
Note that $N_{\rm flux}$ is shown on a logarithmic scale.
The blue data points are solutions with $f^0\neq 0$ or $h^0\neq0$, 
the orange data points are solutions with $f^0=h^0=0$, and
the red curve is the lower bound on $N_{\rm flux}$ 
shown in equation \eqref{bound_01}.
\label{fig_data_01}}
\end{figure}
%%%%%%%%%%%%%%%%%%%%%
%%%%%%%%%%%%%%%%%%%%%
\begin{figure}[p]
\centering
\vskip2em
\includegraphics[width=0.83\textwidth]{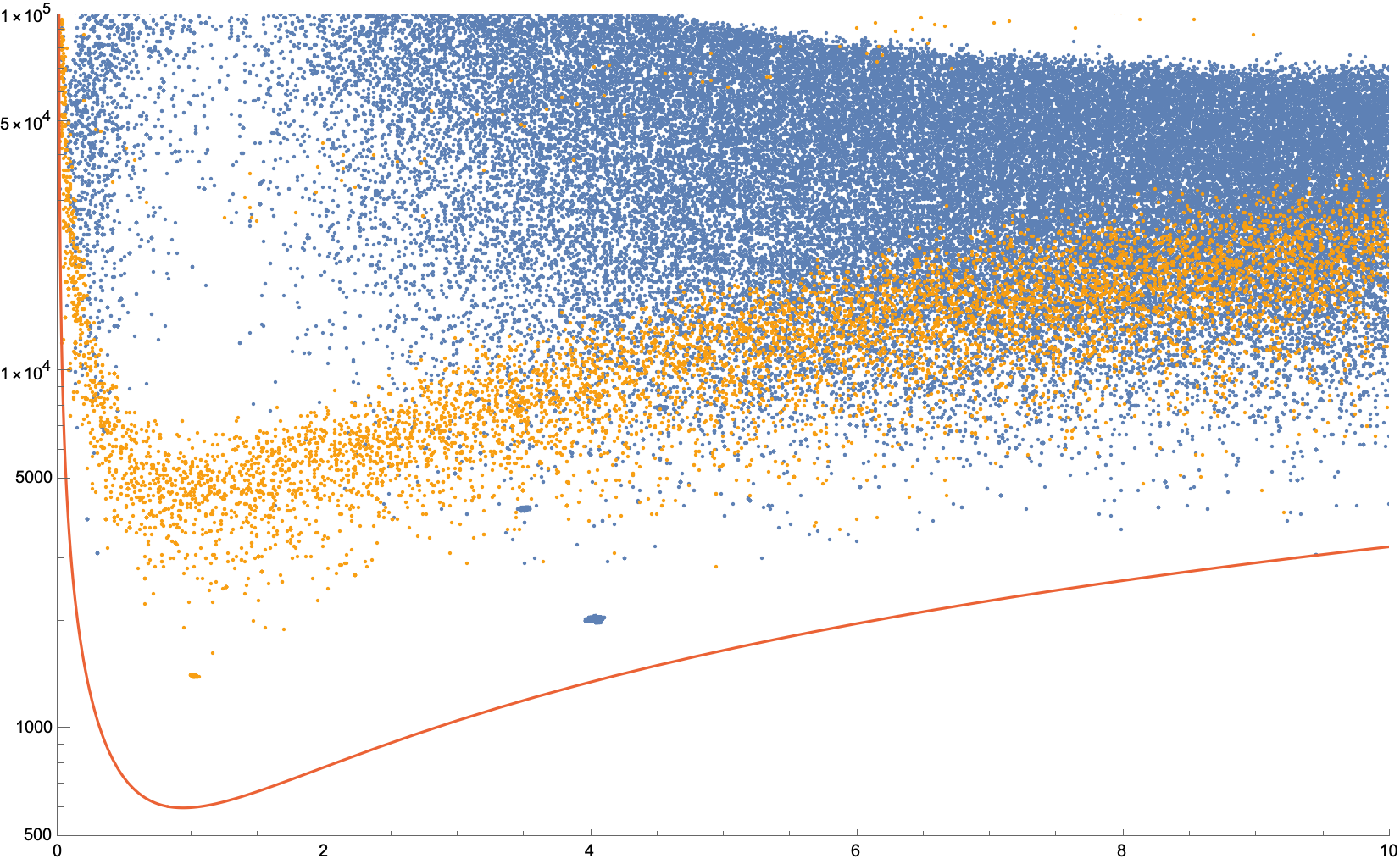}%
\begin{picture}(0,0)
\put(-352,218){\scriptsize$N_{\rm flux}$}
\put(2,4){\scriptsize$s$}
\end{picture}
\caption{Dependence of $N_{\rm flux}$  on 
the dilaton $s$.
Note that $N_{\rm flux}$ is shown on a logarithmic scale.
The blue data points are solutions with $f^0\neq 0$ or $h^0\neq0$, 
the orange data points are solutions with $f^0=h^0=0$, and
the red curve is the lower bound on $N_{\rm flux}$ 
shown in equation \eqref{bound_02}.
\label{fig_data_02}}
\end{figure}
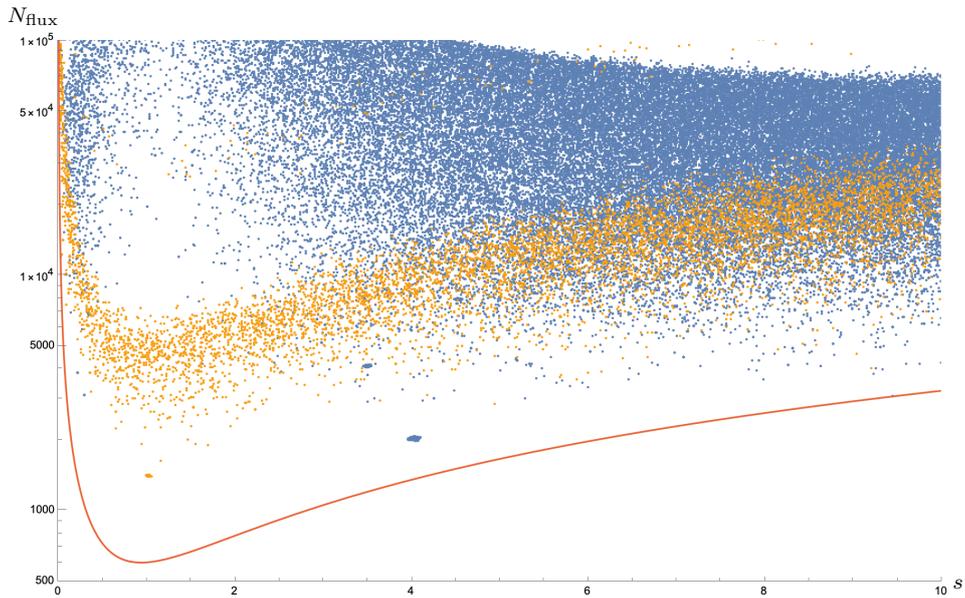
%%%%%%%%%%%%%%%%%%%%%
%%%%%%%%%%%%%%%%%%%%%

\item The minimal flux number also depends  on the value of the dilaton
  $s$.
In figure~\ref{fig_data_02} we have shown $N_{\rm flux}(s)$, 
where we included points in the strong-coupling regime $s\simeq 1$.
We furthermore determined the bound 
\eq{
  \label{bound_02}
  N_{\rm flux}(s)> 321\op s+\frac{282}{s} \,.
}

\item In table~\ref{tab_Nfluxmin} we 
summarized 
the minimal flux number
for a given bound on K\"ahler-cone parameter $\mathsf c$ 
and the dilaton  $s$.
For values $s\lesssim 2$ and $\mathsf c \lesssim 10^{-2}$ we do not believe that 
the corresponding vacua can be trusted
and that corrections 
to the weak-coupling 
and large-complex-structure regime 
have to be taken into account. 
Such corrections are likely to modify these solutions, but
it is beyond the scope of this paper to study this question here.

\item The minimal flux numbers shown in table~\ref{tab_Nfluxmin}
are of the order $N_{\rm flux}\geq\mathcal O(10^4)$ 
and therefore 
exceed the  tadpole bound shown in \eqref{flux_bound}
by about one order of magnitude.
In particular, we did not find any flux choices
that satisfy the tadpole cancellation condition
and stabilize all axio-dilaton and complex-structure moduli.

\item In figures~\ref{fig_data_01} and \ref{fig_data_02} we used two 
different colors to distinguish between flux configurations with $f^0=h^0=0$  
and $f^0\neq 0$ or $h^0\neq0$.
We see that for  $f^0=h^0=0$ 
the minimal flux numbers 
is smaller as compared to flux choices with $f^0\neq 0$ or $h^0\neq0$, which is  in agreement with the discussion in 
\cite{Marchesano:2021gyv}.

\end{itemize}

%%%%%%%%%%%%%%%%%%%%%
%%%%%%%%%%%%%%%%%%%%%
\begin{table}[t]
\begin{equation*}
\arraycolsep8pt
\renewcommand{\arraystretch}{1.3}
\begin{array}{|| l || cccc||}
\hline\hline
& s\geq 1 & s\geq 2 & s\geq 5 & s\geq 10 
\\ \hline\hline
\mathsf c\geq 10^{-3} & 1400 & 1991 & 3023 & 6157 \\
\mathsf c\geq 10^{-2} & 1400 & 1991 & 3023 & 6157 \\
\mathsf c\geq 10^{-1} & 1405 & 1992 & 3385 & 6157 \\
\mathsf c\geq 0.5       & 2993 & 3885 & 4886 & 13218 \\
\mathsf c\geq 1          & 3717 & 5379 & 9345 & 21384 \\
\hline\hline
\end{array}
\end{equation*}
\caption{Smallest values for $N_{\rm flux}$ for given bounds on 
the K\"ahler-cone parameter $\mathsf c$ and the
dilaton  $s$.
\label{tab_Nfluxmin}
}
\end{table}
%%%%%%%%%%%%%%%%%%%%%
%%%%%%%%%%%%%%%%%%%%%

%%%%%%%%%%%%%%%%%%%%%%%%%%%%%%%%%%%%%%%%%%%%%%%
%%%%%%%%%%%%%%%%%%%%%%%%%%%%%%%%%%%%%%%%%%%%%%%

\subsubsection*{Discussion}

Let us now summarize our results  and discuss them in a broader context:
\begin{itemize}

\item Stabilizing the axio-dilaton and complex-structure moduli by numerically minimizing a scalar potential 
becomes very difficult for large moduli-space dimensions. The main problem is to find a suitable 
starting point for the minimization procedure, which cannot be achieved by simple random searches. 
In this paper we presented an algorithm that allows us to find flux vacua 
for large $h^{2,1}_-$ that stabilize all moduli.
To our knowledge, $h^{2,1}_-=50$ is currently the largest moduli-space dimension
for which all axio-dilaton and complex-structure moduli are stabilized by fluxes (without 
imposing additional symmetries).

\item The flux configurations we constructed have a specific structure and are therefore not generic within
the flux space. 
However, our algorithm allows us to find a large number of flux vacua in the large-complex-structure limit
that do not require a tuning of fluxes. In this sense these flux choices are generic (in the 
large-complex-structure regime).

\item The bounds we obtain for the flux number $N_{\rm flux}$ are based on a data set 
with $10^5$ flux vacua. By extending the search time one might 
be able to find vacua with lower $N_{\rm flux}$.

\item The lowest flux numbers we obtain in our search  (c.f.~table~\ref{tab_Nfluxmin}) 
exceed the tadpole bound 
\eqref{flux_bound}
by one order of magnitude. 
Our main conclusion therefore is that constructing consistent flux 
vacua in string theory that stabilize a large number of axio-dilaton and complex-structure moduli is generically difficult.
In particular, we find that the tadpole conjecture \cite{Bena:2020xrh} is satisfied
for our data set.

\end{itemize}

%%%%%%%%%%%%%%%%%%%%%%%%%%%%%%%%%%%%%%%%%%%%%%%
%%%%%%%%%%%%%%%%%%%%%%%%%%%%%%%%%%%%%%%%%%%%%%%

\subsubsection*{Outlook}

We close this section with an outlook for future work:
\begin{itemize}

\item The results presented in this paper provide a starting point for more extensive numerical searches 
for flux vacua with small flux number. Such searches can both employ more computing time 
as well as extending the ranges from which random starting points are drawn
(c.f.~equations \eqref{sol_000} and \eqref{sol_001}).
Our algorithm may also be refined by taking into account further details  
of the minimization condition \eqref{minimum}.

\item In the future we hope to include the non-perturbative contribution to the prepotential \eqref{ppot_03}
for the computation of the minimum condition \eqref{minimum}. 
To determine the non-perturbative part of the prepotential the 
Gopa\-kumar-Vafa invariants are needed which may be provided 
by \texttt{CYTools} in the future.

\item It is important to determine how representative the vacua we found 
are with respect to more general 
flux configurations that stabilize moduli --- not only in the large-complex-structure regime
but also in different regions of moduli space.
This would allow us to revisit arguments on the size of the flux landscape.

\item Solving the tadpole cancellation condition with D7-branes away from the orientifold locus
can, on the one hand, lower the charge $Q_{\rm D3}$ and allow for larger flux numbers
\cite{Collinucci:2008pf}. 
On the other hand, for such D7-branes the NS-NS three-form flux $H_3$ can 
induce a Freed-Witten anomaly \cite{Freed:1999vc} (see \cite{Plauschinn:2018wbo} 
for a brief review).
For such branes one therefore not only has to take into account the interplay between 
fluxes and D3-branes but also between fluxes and D7-branes. 
We are planning to address this question in the future.

\end{itemize}

%%%%%%%%%%%%%%%%%%%%%%%%%%%%%%%%%%%%%%%%%%%%%%%
%%%%%%%%%%%%%%%%%%%%%%%%%%%%%%%%%%%%%%%%%%%%%%%
%%%%%%%%%%%%%%%%%%%%%%%%%%%%%%%%%%%%%%%%%%%%%%%
%%%%%%%%%%%%%%%%%%%%%%%%%%%%%%%%%%%%%%%%%%%%%%%
%%%%%%%%%%%%%%%%%%%%%%%%%%%%%%%%%%%%%%%%%%%%%%%
%%%%%%%%%%%%%%%%%%%%%%%%%%%%%%%%%%%%%%%%%%%%%%%
%%%%%%%%%%%%%%%%%%%%%%%%%%%%%%%%%%%%%%%%%%%%%%%
%%%%%%%%%%%%%%%%%%%%%%%%%%%%%%%%%%%%%%%%%%%%%%%
%%%%%%%%%%%%%%%%%%%%%%%%%%%%%%%%%%%%%%%%%%%%%%%

\vskip3em
\subsubsection*{Acknowledgments}

We thank
Ralph Blumenhagen,
Thomas Grimm,
Umut G\"ursoy,
Damian van de Heisteeg,
Sven Krippendorf,
Jakob Moritz, 
Fabian Ruehle,
Andreas Schachner,
Lorenz Schlechter,
and
Alexandros Singh
for very helpful discussions and communications. 
We furthermore thank Jakob Moritz for providing 
us with Gopakumar-Vafa invariants for our model.
The work of EP is supported by a Heisenberg grant of the
\textit{Deutsche Forschungsgemeinschaft} (DFG, German Research Foundation) 
with project-num\-ber 430285316.

%%%%%%%%%%%%%%%%%%%%%%%%%%%%%%%%%%%%%%%%%%%%%%%
%%%%%%%%%%%%%%%%%%%%%%%%%%%%%%%%%%%%%%%%%%%%%%%
%%%%%%%%%%%%%%%%%%%%%%%%%%%%%%%%%%%%%%%%%%%%%%%
%%%%%%%%%%%%%%%%%%%%%%%%%%%%%%%%%%%%%%%%%%%%%%%
%%%%%%%%%%%%%%%%%%%%%%%%%%%%%%%%%%%%%%%%%%%%%%%
%%%%%%%%%%%%%%%%%%%%%%%%%%%%%%%%%%%%%%%%%%%%%%%
%%%%%%%%%%%%%%%%%%%%%%%%%%%%%%%%%%%%%%%%%%%%%%%
%%%%%%%%%%%%%%%%%%%%%%%%%%%%%%%%%%%%%%%%%%%%%%%
%%%%%%%%%%%%%%%%%%%%%%%%%%%%%%%%%%%%%%%%%%%%%%%
%%%%%%%%%%%%%%%%%%%%%%%%%%%%%%%%%%%%%%%%%%%%%%%

\clearpage
\appendix
\section{Some details on the model}
\label{app_model}

In this appendix we collect some additional information about the model described in 
section~\ref{sec_example}.
\begin{itemize}

\item The polytope in normal form 
from which we construct the 
mirror Calabi-Yau three-fold $\tilde{\mathcal X}$ 
is given by the $6\times 4$ dimensional array
\begin{center}
\begin{minipage}{0.8\textwidth}
\begin{verbatim}
[[ 1  0  0  0] 
 [ 1  3  0  0] 
 [ 1  0  3  0] 
 [-2 -3 -3  0] 
 [ 1  0  0  3] 
 [-2  0  0 -3]] 
\end{verbatim}
\end{minipage}
\end{center}

\item The triangulation 
is performed using  \texttt{triangulate()} of \texttt{CYTools}.
For cross-reference, we note that the points of this triangulation 
can be displayed using \texttt{points()}
which gives in the following $55\times 4$ dimensional array
\begin{center}
\begin{minipage}{0.8\textwidth}
\begin{multicols}{3}
\begin{verbatim}
[[ 0  0  0  0]
 [-2 -3 -3  0]
 [-2  0  0 -3]
 [ 1  0  0  0]
 [ 1  0  0  3]
 [ 1  0  3  0]
 [ 1  3  0  0]
 [-2 -2 -2 -1]
 [-2 -1 -1 -2]
 [-1 -2 -2  0]
 [-1 -2 -2  1]
 [-1  0  1 -2]
 [-1  1  0 -2]
 [ 0 -1 -1  0]
 [ 0 -1 -1  2]
 [ 0  0  2 -1]
 [ 0  2  0 -1]
 [ 1  0  1  0]
 [ 1  0  1  2]
 [ 1  0  2  0]
 [ 1  0  2  1]
 [ 1  1  0  0]
 [ 1  1  0  2]
 [ 1  2  0  0]
 [ 1  2  0  1]
 [-1 -2 -1  0]
 [-1 -1 -2  0]
 [-1  0  0 -2]
 [-1  0  0 -1]
 [ 0 -1  1  0]
 [ 0  0  0 -1]
 [ 0  0  0  1]
 [ 0  1 -1  0]
 [ 1  0  0  1]
 [ 1  0  0  2]
 [ 1  1  2  0]
 [ 1  2  1  0]
 [-1 -1 -1 -1]
 [-1 -1 -1  0]
 [-1 -1  0 -1]
 [-1  0 -1 -1]
 [ 0 -1 -1  1]
 [ 0 -1  0  0]
 [ 0 -1  0  1]
 [ 0  0 -1  0]
 [ 0  0 -1  1]
 [ 0  0  1 -1]
 [ 0  0  1  0]
 [ 0  1  0 -1]
 [ 0  1  0  0]
 [ 0  1  1 -1]
 [ 1  0  1  1]
 [ 1  1  0  1]
 [ 1  1  1  0]
 [ 1  1  1  1]]
 \end{verbatim}
 \end{multicols}
 \end{minipage}
\end{center}

\item The vector $\vec v_*$ introduced above equation \eqref{exp_800} is specified by
\begin{center}
\begin{minipage}{0.8\textwidth}
\begin{verbatim}
[141 -14  13  81  -9  -9  84  31  94  109 -23 -25  49  
  85 -24 -26   1  45  -7  14  46  -7   16  71  71  -8  
   2  15  32  15  32  55 -15 -15  38   53  19  19  65  
  31  48  31  48 -14 -15 -28  21  21  -10  12]
\end{verbatim}
\end{minipage}
\end{center}

\end{itemize}

%%%%%%%%%%%%%%%%%%%%%%%%%%%%%%%%%%%%%%%%%%%%%%%
%%%%%%%%%%%%%%%%%%%%%%%%%%%%%%%%%%%%%%%%%%%%%%%
%%%%%%%%%%%%%%%%%%%%%%%%%%%%%%%%%%%%%%%%%%%%%%%
%%%%%%%%%%%%%%%%%%%%%%%%%%%%%%%%%%%%%%%%%%%%%%%
%%%%%%%%%%%%%%%%%%%%%%%%%%%%%%%%%%%%%%%%%%%%%%%
%%%%%%%%%%%%%%%%%%%%%%%%%%%%%%%%%%%%%%%%%%%%%%%
%%%%%%%%%%%%%%%%%%%%%%%%%%%%%%%%%%%%%%%%%%%%%%%
%%%%%%%%%%%%%%%%%%%%%%%%%%%%%%%%%%%%%%%%%%%%%%%
%%%%%%%%%%%%%%%%%%%%%%%%%%%%%%%%%%%%%%%%%%%%%%%
%%%%%%%%%%%%%%%%%%%%%%%%%%%%%%%%%%%%%%%%%%%%%%%
%%%%%%%%%%%%%%%%%%%%%%%%%%%%%%%%%%%%%%%%%%%%%%%
%%%%%%%%%%%%%%%%%%%%%%%%%%%%%%%%%%%%%%%%%%%%%%%
%%%%%%%%%%%%%%%%%%%%%%%%%%%%%%%%%%%%%%%%%%%%%%%
%%%%%%%%%%%%%%%%%%%%%%%%%%%%%%%%%%%%%%%%%%%%%%%
%%%%%%%%%%%%%%%%%%%%%%%%%%%%%%%%%%%%%%%%%%%%%%%
%%%%%%%%%%%%%%%%%%%%%%%%%%%%%%%%%%%%%%%%%%%%%%%
%%%%%%%%%%%%%%%%%%%%%%%%%%%%%%%%%%%%%%%%%%%%%%%
%%%%%%%%%%%%%%%%%%%%%%%%%%%%%%%%%%%%%%%%%%%%%%%

\clearpage
\nocite{*}
\bibliography{references}
\bibliographystyle{utphys}

%%%%%%%%%%%%%%%%%%%%%%%%%%%%%%%%%%%%%%%%%%%%%%%
%%%%%%%%%%%%%%%%%%%%%%%%%%%%%%%%%%%%%%%%%%%%%%%
%%%%%%%%%%%%%%%%%%%%%%%%%%%%%%%%%%%%%%%%%%%%%%%
%%%%%%%%%%%%%%%%%%%%%%%%%%%%%%%%%%%%%%%%%%%%%%%
%%%%%%%%%%%%%%%%%%%%%%%%%%%%%%%%%%%%%%%%%%%%%%%

\end{document}